\documentstyle[12pt,aps,epsf]{revtex}
\begin{document}
\baselineskip 0.25in
\title{Noise in an insect outbreak model}
\author{Bao-Quan  AI$^{1}$, Wei  CHEN$^{2}$ , Xian-Ju
WANG$^{1}$,  Guo-Tao  LIU$^{1}$,\\De-Hua Wen$^{1,3}$, Hui-Zhang Xie$^{3}$and Liang-Gang
LIU$^{1}$}
\maketitle
\begin{center}
{\sl$^{1}$Department of Physics, ZhongShan University,
GuangZhou, China\\
$^{2}$Department of Physics, JiNan  University, GuangZhou,
China.\\
$^{3}$Department of Physics, South China University of
technology,\\GuangZhou, China.\\}
\end{center}
\parskip 0pt

\begin{abstract}
\baselineskip 0.25in We study the steady state properties of an
insect (spruce budworm) outbreak model in the presence of Gaussian
white noise. Based on the corresponding Fokker-Planck equation the
steady state solution of the probability distribution function and
its extrema have been investigated. It was found that fluctuations
of the insect birth rate reduces the population of the insects
while fluctuations of predation rate and the noise correlation can
prevent the population of the insects from going into extinction.
Noise in the
model can induce a phase transition. \\
Pacs numbers: 87. 10. +e, 05. 40. -a, 02. 50. Ey.
\end{abstract}
\vskip 0.2in
 \baselineskip 0.25in
\section {Introduction}
\parskip 0pt
\indent Recently, Nonlinear dynamical systems with random noise
have been paid much attention both theoretically and
experimentally. Phenomena such as noise-induced transitions,
stochastic resonance, resonant activation, noise-induced spatial
patterns are a few examples of extensive investigations
\cite{1}\cite{2}\cite{3}. In most of these areas the noise affects
the dynamics through system variables, i.e., the noise is
multiplicative in nature \cite{4,5}. The focal theme of these
types investigation is the steady state properties of systems in
which the fluctuations, generally applied from outside, are
considered, independent of the system's characteristic
dissipation.
 On the level of a Langevin-type description of a dynamical system,
the presence of noise can change the dynamics of the system
\cite{6,7}.  Noise processes have found applications in a broad
range of studies, such as the steady state properties of a single
mode laser \cite{8}, bistable kinetics \cite{9}, directed motion
in spatially symmetric periodic potentials \cite{10}, stochastic
resonance in linear systems \cite{11}, and steady state entropy
production \cite{12}. In this paper we study an insect outbreak
model in the presence of the correlated noise, and show
how noise can dynamically affect the ecosystem. \\
\section {An Insect Outbreak Model: Spruce Budworm}
\indent A practical model which exhibits two positive linearly
steady state populations is that for the spruce budworm, which
can, with ferocous efficiency, defoliate the balsam fir. They are
major problem in Canada \cite{13}, Ludwing \cite{14} (1978)
considered the budworm population dynamics to be governed by the equation\\
\begin{equation}\label{e1}
\frac{dN}{dt}=r_{B}N(1-\frac{N}{k_{B}})-P(N).
\end{equation}
\indent Here $r_{B}$ is the linear birth rate of the budworm and
$K_{B}$ is carrying capacity which is related to the density of
the foliage available on trees. The $P(N)$ term represents
predation, generated by birds. We take the form for $P(N)$
suggested by Ludwing (1978), namely $BN^{2}/(A^{2}+N^{2})$, where
$A$ is  a positive constant and $B$ represents the predation rate
of the birds. The dynamics of $N(t)$ is
then governed by\\
\begin{equation}\label{e2}
\frac{dN}{dt}=r_{B}N(1-\frac{N}{k_{B}})-\frac{BN^{2}}{A^{2}+N^{2}}.
\end{equation}
\indent Before analysing  the model we will express it in
nondimensional terms. Here we introduce nondimensional quantities
by\\
\begin{equation}\label{e3}
x=\frac{N}{A},
r=Ar_{B},q=\frac{K_{B}}{A},\tau=\frac{t}{A},\beta=B.
\end{equation}
Upon on substitution (2) becomes\\
\begin{equation}\label{4}
  \frac{dx}{d\tau}=rx(1-\frac{x}{q})-\frac{\beta x^{2}}{1+x^{2}}.
\end{equation}
\indent Now, if some environmental external disturbances make both
the birth rate and the predation rate fluctuate, it is likely to
affect $r$ and $\beta$ in the form of multiplicative noises that
are connected through a correlation parameter $\lambda$. As a
result we have
\begin{equation}\label{5}
\frac{dx}{d\tau}=rx(1-\frac{x}{q})-\frac{\beta
x^{2}}{1+x^{2}}+x(1-\frac{x}{q})\Gamma(t)-\frac{x^{2}}{1+x^{2}}\xi(t).
\end{equation}
Where $\Gamma(t)$ and $\xi(t)$are Gaussian white noises
with the following properties:\\
\begin{equation}\label{6}
\langle \Gamma(t)\rangle=\langle \xi(t)\rangle=0,
\end{equation}
\begin{equation}\label{7}
\langle\Gamma(t)\Gamma(s)\rangle=2D\delta(t-s),
\end{equation}
\begin{equation}\label{8}
\langle\xi(t)\xi(s)\rangle=2\sigma\delta(t-s),
\end{equation}
\begin{equation}\label{9}
\langle\xi(t)\Gamma(s)\rangle=\langle\Gamma(t)\xi(s)\rangle=2\lambda\sqrt{D\sigma}\delta(t-s).
\end{equation}
\indent Where $D$ and $\sigma$ are the strength of noise
$\Gamma(t)$ and $\xi(t)$, respectively, and $\lambda$ denotes the
degree of correlation between $\Gamma(t)$ and $\xi(t)$ with
$0\leq\lambda\leq 1$.\\
\indent We can derive the corresponding Fokker-planck equation for
evolution of Steady Probability Distribution (SPDF) based on Eq.
(5)-Eq.(9). The equation is  as follows \cite{15}
\begin{equation}\label{10}
\frac{\partial P(x,t) }{\partial t}=-\frac{\partial
A(x)P(x,t)}{\partial x}+\frac{\partial^{2}B(x)P(x,t)}{\partial
x^{2}},
\end{equation}
where
\begin{equation}\label{11}
A(x)=h(x)+Dg_{1}(x)g_{1}^{'}(x)+\lambda\sqrt{D\sigma}g_{1}(x)g_{2}^{'}(x)+\lambda\sqrt{D\sigma}g_{1}^{'}(x)g_{2}(x)
+\sigma g_{2}(x)g_{2}^{'}(x),
\end{equation}
\begin{equation}\label{12}
B(x)=Dg_{1}^{2}(x)+2\lambda\sqrt{D\sigma}g_{1}(x)g_{2}(x)+\sigma
g_{2}^{2}(x).
\end{equation}
\indent Here $h(x)=rx(1-\frac{x}{q})-\frac{\beta x^{2}}{1+x^{2}}$,
$g_{1}(x)=x(1-x/q)$, $g_{2}(x)=-\frac{x^{2}}{1+x^{2}}$. The steady
probability distribution of the Fokker-Planck equation is given by
\cite{15}
\begin{equation}\label{e15}
 P_{st}(x)={N_{0}\over
 B(x)}\exp[\int^{x}\frac{A(x^{'})}{B(x^{'})}dx^{'}].
\end{equation}
 \indent Where $N_{0}$ is the normalization constant.

\section{The fluctuation of birth rate in the model}
\indent If we only consider the fluctuation of birth rate on the
model, namely $\sigma=0$ and $\lambda=0$, we can get
\begin{equation}\label{e10}
A(x)=rx(1-\frac{x}{q})-\frac{\beta
x^{2}}{1+x^{2}}+Dx(1-\frac{x}{q})(1-\frac{2x}{q}),
\end{equation}
\begin{equation}\label{e11}
  B(x)=D[x(1-\frac{x}{q})]^{2}.
\end{equation}
 From Eq. (13), using the forms of $A(x)$ and $B(x)$, we get the  following integral forms
of the SPDF \cite{16}.
\begin{equation}\label{e15}
P_{st}(x)=\frac{N_{0}}{|g_{1}(x)|}\exp[\frac{f(x)}{D}].
\end{equation}
 Here \\
\begin{equation}\label{e16}
  g_{1}(x)=x(1-\frac{x}{q}),
  \end{equation}
\begin{equation}\label{e17}
f(x)=r\ln |\frac{x}{x-q}|+\frac{\beta q^{2}}{1+q^{2}}(\arctan
x+\frac{1}{x-q})+\frac{2\beta q^{3}}{(1+q^{2})^{2}}(\ln
|\frac{x-q}{\sqrt{1+x^{2}}}|-q\arctan x).
\end{equation}
\indent The extrema of the SPDF are calculated using the condition
$A(x)-B^{'}(x)=0$\\
\begin{equation}\label{e18}
r(1-\frac{x}{q})-\frac{\beta}{1+x^{2}}-D(1-\frac{x}{q})(1-\frac{2x}{q})=0.
\end{equation}
\indent  As for the parameters of the above equations we adopt
$r=1.0,q=10.0,\beta=2.0$,
 The results are represented in Fig.1-Fig.2\\

\begin{figure}[htb]
\centerline{\epsfxsize 10cm \epsffile{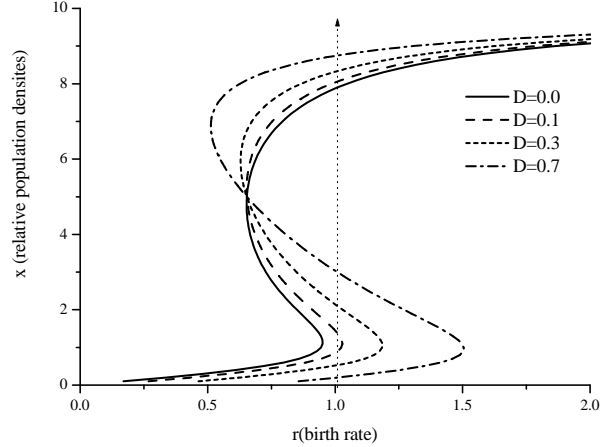}} \label{fig1}
\caption{Plot of the extrema of SPDF as a function of $x$ for
different noise strength  values: $D=0.0, 0.1, 0.3,0.7$,using
$\beta=2.0$, and $q=10.0$.}
\end{figure}

\indent In  Fig.1, for zero noise strength, the curve at $r=1.0$
gives only one $x$ value, but as the value of $D$ increases, the
curve changes. The curve shows three $x$ values at $r=1.0$. From
the figure we can see that the noise strength
can change the state of the system.\\

\indent  In Fig.2 we show the effect of the noise strength $D$ on
the SPDF. For a small value of $D$, the SPDF shows a single peak
region, which changes for larger values of $D$ (see Fig.2). As the
value of $D$ increases, the peak for the larger values of $x$
decreases. At the same time, for small values of $x$, a new peak
appears. It can be said that the noise causes the system become
from two states to one state, namely noise can induce a phase
transition. On the other hand, since $x$, $D$ give the relative
budworm population and the fluctuation of the budworm's birth
rate, respectively, we can see from the figure the effect the
fluctuation of the birth rate makes on the growth of budworms,
they can even make the budworms go into extinction.\\
\begin{figure}[htb]
\centerline{\epsfxsize 10cm \epsffile{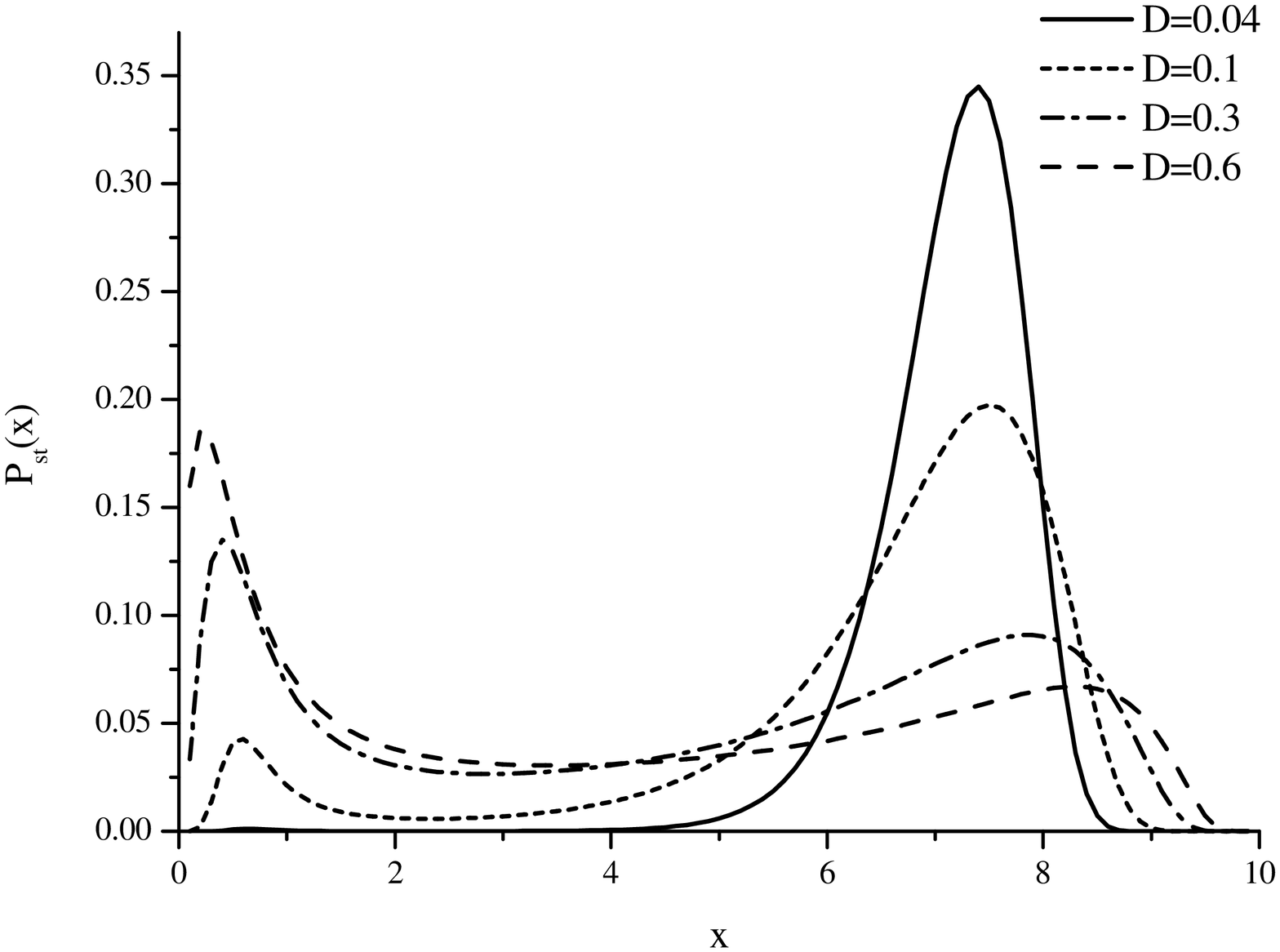}}\label{fig2}
\caption{Plot of $P_{st}(x)$(denotes the probability) against $x$
for different  noise strength values: $D=0.04, 0.1, 0.3 ,0.6$,
using $r=1.0,\beta=2.0$, and $q=10.0$.}
\end{figure}

\section{The fluctuations of the predation rate in the model}
\indent On the other hand, if only the fluctuation of predation
rate is investigated, namely $D=0$ and $\lambda=0$ we can get the
stationary
probability distribution function similar to Eq.(14)\\
\begin{equation}\label{e21}
P_{st}(x)=\frac{N_{0}}{|g_{2}(x)|}\exp[\frac{f(x)}{\sigma}],
\end{equation}
\indent Where \\
\begin{equation}\label{e22}
g_{2}(x)=\frac{x^{2}}{1+x^{2}},
\end{equation}
\begin{equation}\label{e23}
f(x)=-\frac{r}{3q}x^{3}+\frac{r}{2}x^{2}-(\beta+\frac{2r}{q})x+2r\ln
x +(\beta+\frac{r}{q})x^{-1}-\frac{r}{2}x^{-2}.
\end{equation}
In order to discuss the effect of the  fluctuation of the
predation rate on the system we adopt $r=1.0,q=10.0,\beta=2.26$,
the results are shown in Fig.3.\\

\indent Fig. 3  shows the effect of the noise strength $\sigma$ on
the SPDF. For a small value of $\sigma$, the SPDF shows the
typical bistable region (see Fig.3) which vanishes for large
values of $\sigma$. As the value of $\sigma$ increases the peak at
the small $x$ value decreases drastically, while for a large
$\sigma$ value
 the curve has a single peak at the large $x$ value . Since
$x$ denotes the budworm relative population, it is clear from
Fig.3 that, with an increase in the fluctuations of the predation
rate, the budworm recover from going into extinction.\\
\begin{figure}[htb]
\centerline{\epsfxsize 10cm \epsffile{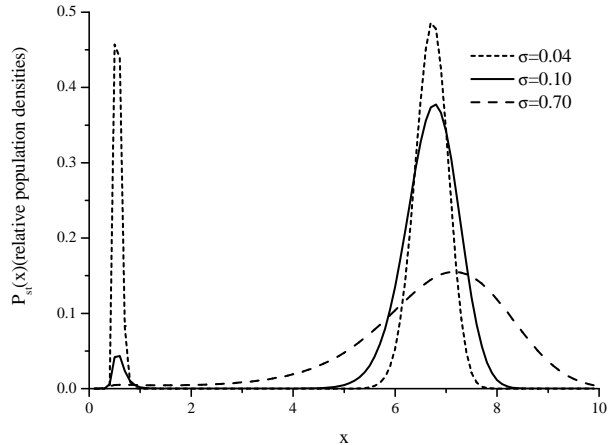}}\label{fig3}
\caption{Plot of $P_{st}(x)$(denotes the probability) against $x$
for different noise strength  values: $\sigma=0.04, 0.10, 0.70$
using $r=1.0, \beta=2.26$, and $q=10.0$.}
\end{figure}
\section{Effect of noise correlation in the system}
\indent Since the two type of fluctuations have the common origin,
(environmental external disturbance), we will consider the effect
of the correlation between the birth rate fluctuation and the
predation rate fluctuation. Based on Eq. (10)-Eq. (13) the
stationary probabilities distribution function is represented by
Fig. 4 with $r=1.0$, $\beta=2.0$, $D=0.3$, $\sigma=0.3$.

\indent Fig. 4 gives the effect of noise correlation parameter
$\lambda$ on SPDF. For a small $\lambda$ value the SPDF shows a
typical bistable region which will vanish for the large value of
$\lambda$. As the correlation parameter increases the height of
the peak on the small $x$ value decreases, while the height of the
peak on the large $x$ value increases, namely probabilities flow
from the small $x$ value to the large $x$ value. Since $x$ denotes
the budworm relative population, it is clear from Fig. 4 that the
correlation of the noises is a advantageous factor for the growth
of the budworm.
\begin{figure}[htb]
\centerline{\epsfxsize 10cm \epsffile{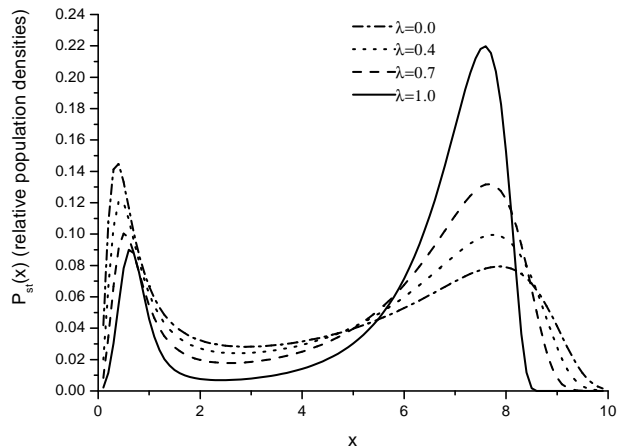}}\label{fig4}
\caption{Plot of $P_{st}(x)$(denotes the probability) against $x$
for different   correlation parameter values: $\lambda=0, 0.4,
0.70, 1.0$ using $r=1.0, \beta=2.0$ ,$q=10.0$, $D=0.3$, and
$\sigma=0.3$.}
\end{figure}
\section {Summary}
\indent In this paper, we studied the steady state properties of
an insect outbreak model in the presence of the correlated noise.
Birth rate fluctuation, predation rate fluctuation and the
correlation of the noises are investigated. Birth rate fluctuation
work against the growth of the budworm population, while both
fluctuation of the predation rate and the correlation of the
noises can  prevent the budworm population from going into
extinction. On the other hand, noise can dynamically affect the
ecosystem. The noise from the birth rate fluctuation can make the
system change from a single steady state to a bistable state,
while noise from the predation rate fluctuation induces the system
change from a bistable state to a single steady state. Thus in the
insect outbreak model, noise from the fluctuation of the
parameters can induce a phase transition. This viewpoint is
completely novel in the traditional viewpoint it is  the
stochastic force that disturbs the
phase transition.\\

{\bf Acknowledgements}\\
 \indent The project supported by National
Natural Science Foundation of China (Grant No. of 10275099) and
GuangDong Provincial Natural Science Foundation (Grant No. of
021707 and 001182).\\

\end{document}